\documentclass{elsart}

 \usepackage{epsfig}

\usepackage{amssymb}

\begin{document}

\begin{frontmatter}



\title{Influence of thermal history on the structure and properties
 of silicate glasses}


\author[label1]{C. Levelut, R. Le Parc, A. Faivre}

\address[label1]{Laboratoire des Collo\"{\i}des, Verres 
et Nanomat\'eriaux, CNRS UMR 5587, Universit\'e Montpellier II, France}

\author[label2]{B. Champagnon}
\address[label2]{Laboratoire de Physico-Chimie des Mat\'eriaux 
Luminescents, CNRS/UMR 5620, Universit\'e Lyon 1, France}

\begin{abstract}

We studied a set of float glass samples
prepared with different fictive temperature 
by previous annealing around the glass transition temperature. 
We compared the results to previous measurements on a series of amorphous silica samples, 
also prepared with different fictive temperature.
We showed that the 
 modifications on the structure at a local scale are very small, 
the changes of physical 
properties are moderate but 
the  changes on density fluctuations at a nanometer scale are 
rather large: 12 and 20\% in float 
glass and silica, for relative changes of fictive temperature equal to 
13 and 25\% respectively. 
 Local order and mechanical properties of silica vary in the
 opposite way compared to  float glass (anomalous behavior)
 but the  density fluctuations  in both glasses increase with  
temperature and fictive temperature.

\end{abstract}

\end{frontmatter}

\section{Introduction}
\label{intro}
The structure and formation of glass, though being studied 
for decades, are still  puzzling problems. Liquids  quenched with
 different cooling rate, or having different thermal  histories, provide
 glasses with different  structures and properties \cite{Tool}.
The local structure of silicate glasses is rather well-known: 
 the ``building units'' 
are similar to that of the crystalline counterparts (SiO$_4$ tetrahedra), 
but at intermediate range the tetrahedra are linked together in a 
disordered fashion with a
bound angle distribution, which has been shown 
to depend on thermal history in silica \cite{Tomazawa2005}.
 Modifiers, 
such as alkaline species,   induce breaking of 
intertetrahedral bounds. This phenomenon is often referred to as
 ``depolymerization'' of the glassy network. 
  No much is known about the structure at larger
 scale. However density and concentration
fluctuations present in the liquid are expected to take place
 at the nanometer scale
and to be partly frozen-in around the glass transition.

 In order to determine the  
structural changes on different length
 scales and how the macroscopic properties are affected by thermal history, 
we had studied silica samples (glass transition \mbox{$T_g=1260$ $^{\circ}$C})
 prepared with well defined thermal history    
  \cite{Champagnon2000,LeParcChampagnon2002,ChampagnonLeParc2002,LevelutFaivre2002,LeParc2005}. In this paper,
 we present new results about a commercial  ``float'' glass 
(\mbox{$T_g=560$ $^{\circ}$C}), 
very widespread because used as window glass. Both composition are examples of usual silicate glasses, 
among the most technologically important. 
All samples have  well 
characterized thermal histories, prepared by thermal annealing at different temperatures  in 
the glass transition range 
 and then quenching on a metallic plate.  Samples prepared this way 
keep the structural organization  
present in the liquid at the  temperature at which the 
structure is frozen-in. This temperature, called fictive temperature, 
$T_f$, \cite{Tool}, 
can be used as a characteristic of the   simple  thermal histories (stabilization and quenching) performed in this work.
The fictive temperature of the samples is
equal to the annealing temperature if the annealing time is long 
enough to stabilize the structure and 
properties (longer than the relaxation 
time for viscosity, $\frac{\eta}{G_{\infty}}$, where $\eta$ is 
the viscosity 
 and $G_{\infty}$ the infinite frequency shear modulus)
and the quench fast enough  to retain its configuration.

\section{Experimental Procedure}

 The commercial ``float'' glass, 
of composition  72\% SiO$_2$, 14\% Na$_2$O, 
9\% CaO, 3\% MgO, 1\% Al$_2$O$_3$ plus other minor oxides, is
 provided by 
Saint-Gobain Recherche, Aubervilliers, France. Four heat treatments 
 have been performed, in order to prepare samples with fictive 
temperature $T_f$ ranging 
from 480 $^{\circ}$C  to 
\mbox{620 $^{\circ}$C}   
annealing. 
The annealing times range from  2h for \mbox{620 $^{\circ}$C} to 16 weeks for \mbox{480 $^{\circ}$C}. They were
estimated from the stabilization  time for silica glass as equivalent viscosity.  Moreover,  we also
measured the SAXS signal on a series of samples annealed for 
 different times at  \mbox{480 $^{\circ}$C} and also during some in situ annealing, in order to check that 
 stabilization was reached. It should be noted that temperature dependent SAXS measurements
on synchrotron were  performed on a 12 week annealed sample but the results showed that this time was
insufficient, and thus 16 week treatment were performed later.
The results are compared to measurement in 
 a very high purity (low OH content) type I  \cite{Bruckner1970} 
silica glass.
 The heat treatments performed to prepare different samples
 with different thermal histories,
corresponding to fictive temperatures, $T_f$, ranging from 
\mbox{1100 $^{\circ}$C} to 
\mbox{1500 $^{\circ}$C} have been  described previously
\cite{LeParcChampagnon2002,LevelutFaivre2002,LevelutFaivre2005}.
For silica, the spectroscopic
measurements yield a method to check that the fictive temperature
 is most probably equal to the annealing temperature.

The local structure  is probed  using in situ Raman spectroscopy as 
a function of  temperature for float glass, 
with an incident wavelength \mbox {$\lambda= 5145$ \AA},
using a Jobin-Yvon T64000 set-up.
Raman room temperature  measurements for silica have been reported
 before \cite{Champagnon2000,LeparcChampagnon2001}.
Larger scale structure 
 fluctuations at the nanometer scale)
is investigated up to \mbox{700 $^{\circ}$C}
 (\mbox{1500 $^{\circ}$C} for silica) using Small 
Angle X-ray Scattering (SAXS) on D2AM  instrument 
at ESRF (Grenoble, France), with an incident energy
 E=18keV, using a molybdenum furnace \cite{Soldo1998}.
 Finally,
 we determined  temperature dependent
macroscopic longitudinal  sound velocity (or  modulus), using
  Brillouin scattering (\mbox{$\lambda=5145$ \AA}) performed with
 a high resolution spectrometer \cite{Vacherspectro,Vacherspectro2} 
in backscattering geometry. All the optical 
temperature dependent measurements where achieved  
using  a Linkam TS1500 device.

\section{Results}

Results about silica have been described before and we will first present 
data obtained in float glass.
Raman spectra of float glass  (Insert of Fig. \ref{vavram}) exhibit
 two main features: a broad peak around 
\mbox{600 cm$^{-1}$}, and a line around \mbox{1100 cm$^{-1}$}. 
The position of
 the Raman line around 1100 cm$^{-1}$ 
 is shifted to lower frequency 
when the temperature increases (Fig. \ref{vavram}a), with a  change 
of slope at the 
glass transition (around \mbox{560 $^{\circ}$C}).  At room temperature,
 the line around \mbox{1100 cm$^{-1}$} 
 is shifted down by  about 
\mbox {3 cm$^{-1}$ } {\it i.e.} 0.3\%   for a fictive temperature change, 
$\Delta T_f/T_f=17\%$ (Fig \ref{vavram}b). We also observed 
around the glass transition, a
 (rather small) relaxation of the sample stabilized 
at  \mbox{620 $^{\circ}$C} (Fig. \ref{vavram}a).
The position of the 1100 cm$^{-1}$ line evolves toward that of the 
sample of lowest fictive temperature as
  the relaxation time of this sample is shorter than the measuring 
time (10 min): the sample lowers its $T_f$ during the measurements.

 The amplitude of the density (concentration) fluctuations can be 
deduced from the SAXS spectra by extrapolating 
to $q$=0 the signal just below the first diffraction peak (observed 
around \mbox{$q=1.5$ \AA$^{-1}$} for float glass) (Fig. \ref{vavsaxs}a).
Like in silica glass \cite{LevelutFaivre2002}, we  observe  
 a rather large influence of $T_f$ on the amplitude 
of fluctuations at room temperature (Fig. \ref{vavsaxs}b).  
  The sample treated 12 weeks at \mbox{$T_f=480 ^{\circ}$C} 
 is probably not 
fully stabilized. Thus the  fictive temperature range to consider 
for those SAXS measurements is only
\mbox{510-620 $^{\circ}$C}.  The SAXS intensity vary then 
by about  12\% for a fictive temperature change of 
13\%. We observe, like for the Raman shift, a clear relaxation 
in the glass transition region for the sample with
 the highest fictive temperature. 
 On the other hand for the two samples treated at the
 lowest temperatures (480 and 510 $^{\circ}$C), we observe a ``delay''
 to reach the supercooled liquid curve when heating, due to
  relaxation times longer than typical experimental times.

Finally, the influence of temperature (Fig. \ref{vavbril}a) 
and fictive temperature (Fig. \ref{vavbril}b) on sound velocity 
of float glass
has been measured by Brillouin scattering. The effect is  very similar
 to that on the 
SAXS intensity $I(q=0)$ except that the sound velocity decreases 
when both $T$ and $T_f$ increases.  The sample
 stabilized at \mbox{480 $^{\circ}$C} has been 
heat-treated 16 weeks (as for Raman measurements) and can be considered 
as fully stabilized. We observed
the influence of fictive temperature in the glassy state (a variation 
of 2.45\% for a 
 fictive temperature change of 16\%) as well as a relaxation effect
 around 
the glass transition  for the  samples of $T_f=620$ and
 \mbox{590 $^{\circ}$C}. A ``delay'' effect can be noticed for
 $T_f=480$ and \mbox{510 $^{\circ}$C}. 
 The changes are clearly much lower on the 
sound velocity than on the SAXS intensity.

In silica, vibrational spectroscopy measurements 
showed that when $T_f$ increases, the frequency  
of the main band at 440 cm$^{-1}$, related to symmetric 
 $\widehat{\mbox{Si-O-Si}}$ bend  
increases by 2\%.  
\cite{ChampagnonLeParc2002}. 
The area of  defect lines (at 480  and \mbox{520  cm$^{-1}$}) also
 increases with increasing $T_f$.
On the other hand, the  changes on the amplitude of density 
fluctuations,
 $I(q=0)$, have been reported to 
be very large below the glass transition, about 20\% at room temperature
 for a $\Delta T_f/T_f$=25\% \cite{LevelutFaivre2002,LevelutFaivre2005}. 
A structural relaxation effect (decrease of $I(q=0)$ for high $T_f$)
 has  also been observed  in the glass transition range 
\cite{LevelutFaivre2002}.
 In situ isothermal relaxation can also be observed   
 as a function of time (fig. \ref{silicerelax}). A typical 
relaxation time for
 as-received sample annealed at \mbox{1175 $^{\circ}$ C}
is about 40 min, 
comparable to the evolution time observed by Raman and infrared
 spectroscopies for samples annealed for
different times at a given temperature (30 to 70 min) \cite{Leparcthese}. 
The sound velocity vs temperature curves in silica, deduced 
from Brillouin scattering, are very different 
from that of  float glass: the sound velocity increases with 
temperature, and no change of slope 
is observed at the glass transition \cite{LeParc2005}. 
The   influence of fictive temperature in silica is very small,
 the sample of lower fictive temperature being softer (having 
a lower sound velocity) by 0.2\% for $\Delta T_f=25\%$ \cite{LeParc2005}.

\section{Discussion}

The changes of Raman 
 spectra 
in silica with $T_f$ can be interpreted as 
a decrease of the average angle $\widehat{\mbox{Si-O-Si}}$ 
\cite{Tomazawa2005,Agarwal,Galeener1983,ChampagnonLeParc2002}, 
and thus as a local densification. On the other hand the change on
 amplitude of the so-called 
defect lines  as a function of $T_f$ also
 indicates that either the number of  small cycles, involving 3
 to 4 SiO$_4$ tetrahedra,
increases with $T_f$, either those cycles are more decoupled
 from the other vibrations. 
The first hypothesis also implies local densification and 
decrease of $\widehat{\mbox{Si-O-Si}}$ angle
 (\mbox{$d\theta/dT_f=-0.00623 ^{\circ}/^{\circ}$C}) \cite{Tomazawa2005}.
In Raman spectra of sodium silicate glasses, the  feature  
at 1100 cm$^{-1}$  is 
attributed  to vibrations of different $Q_n$ species
with a main contribution from $Q_3$,  a low frequency side 
due to $Q_2$ and a high frequency 
side due to $Q_4$ \cite{Zotov1999} ($Q_n$ is a SiO$_4$ 
tetrahedron linked to $n$ other tetrahedra toward $n$ 
bridging oxygens). The Raman spectra 
of float glass being very similar to that of sodium 
silicate glass, we infer that a  shift downward can be interpreted 
as an increase of the Q$_2$ population. This result is 
in agreement with a local depolymerization of the silicate 
network  with increasing fictive temperature, 
also observed by RMN in soda-lime silicate glass \cite{Brandis}.  
Almost no study of the  
 influence 
 of the heat-treatment on the Raman spectra of 
float glass or other silicate glass can be found 
in the literature \cite{Tan}. Few 
studies of the influence of $T_f$
have been carried out by infrared spectrometry 
 in float glass \cite{Agarwal} and in a  lithium 
aluminosilicate glass \cite{TomazawaSakamoto2003}. 
The changes induced  in the infrared spectrum by modifying
 $T_f$   are difficult to observe due to broad features in the 
spectra. The frequency changes in float 
glass is in opposite sense compared to what is observed in infrared
 spectra of silica. Surprisingly,
in  the aluminosilicate glass with 70\% of SiO$_2$, 15\%Al$_2$O$_3$, 
15\%Li$_2$O, the frequency of infrared 
feature at 1080 cm$^{-1}$
 varies in the same way as it does in silica \cite{TomazawaSakamoto2003},
 thus showing the complexity 
of behavior in silicate glass, probably due to several
 contributions being mixed in one single peak. 
The Raman spectra are also 
difficult to interpret in aluminosilicate glasses because
 of the interplay of several contributions (for
 example  modes involving Si and Al are expected to have 
very similar frequencies) which may  vary in opposite sense.

For both silica and float glasses, a $T_f$ increase
 induces a rather noticeable increase 
of the amplitude of density fluctuations. Moreover the evolution
 with temperature is similar 
for silica glass and float glass: an increase of  $I(q=0)$  with
 temperature 
and a change of slope, toward larger slope at the glass transition. 
The amplitude of 
density fluctuation is larger in silica, frozen-in at 
high temperature, than in float  glass. 
The ratio of extrapolated scattered intensity, respectively around 
22 and 12 e.u/molecules for 
silica and float glass, is roughly equal to the ratio of glass 
transition temperatures : 1533 
and \mbox{833 K}. 
It can be noted also that
 the observation of the relaxation  around  the glass transition in
 silica glass is a rather new 
result. Indeed, such relaxation effect is usually observed in silicates 
glasses by dilatometry 
measurements  or differential scanning calorimetry. However, in 
silica both kinds of measurements 
are difficult due to the high $T_g$
and also to the low thermal expansion of silica.

 The  sound velocity vary in the opposite way for  silica and float
 glass, and the influence of $T_f$ is very low (0.4\%) in silica.
Both effects are due to the  elastic anomaly of silica.

\section{Conclusion}
For both silica and float glass samples, we observed that the 
modifications induced 
  on the local order, by changes of the fictive temperature, as 
probed by spectroscopic methods 
 are small. A noticeable influence of fictive temperature is 
observed on the  macroscopic elastic modulus
of float glass. 
In silica, the influence of $T_f$ on 
elastic modulus is  small and is in opposite sense compared to 
float glass 
due to the inelastic property anomalies of silica. 
SAXS measurements show that, for both glasses, the higher fictive 
temperature 
contains fluctuations of larger amplitude, and the effect is rather
 important: changes of more than  10\%
in the SAXS intensity are observed  for fictive temperature changes
 of the same order.
 This result could be considered as rather intuitive but does not 
correspond to previous observation in amorphous polymers where the 
effect of thermal history is hardly observed 
\cite{Curro1984,Wendorff1973}.
\section*{Acknowledgment} We thank S. Abensour,  Saint-Gobain
 Recherche, Aubervilliers, who provide the float glass sample.
We wish to thank O. Geyamond, S. Arnaud,  B. Caillot, and 
J.-F B\'erar (Laboratoire de Cristallographie, Grenoble),
 J.-P. Simon (Laboratoire de Thermodynamique et Physicochimie
  M\'etallurgiques, Grenoble) and L. David (GEMPPM, Lyon)
 for assistance in using beam-line BM02. We also thanks 
ESRF staff for operating the synchrotron radiation facilities
 and J.-L. Hazemann (Laboratoire de Cristallographie)
 for supplying the  furnace.We thanks G. Garcia for his participation in preparation of samples and
 Brillouin measurements for the float glass.

\newpage
\centerline{Figure captions}

\begin{figure}

\caption{a) (left) Evolution with temperature 
of the Raman line around \mbox{1100 cm$^{-1}$}, for two samples 
corresponding to fictive temperatures 480 and \mbox{620$^{\circ}$ C}, 
respectively. b)(right) Position of the line around 
\mbox{1100 cm$^{-1}$} at room temperature as a function of $T_f$. Insert:  Example of Raman spectra in float glass \label{vavram}}
\end{figure}

\begin{figure}
 
 \caption{a)SAXS extrapolated intensity  as a function of temperature 
for 4 float glasses with different heat treatment. The sample annealed 
12 week at \mbox{480 $^{\circ}$C} is not fully stabilized. The error on $I(q=0)$ is $\pm 2\%$ b) I(q=0) at room temperature as a function of $T_f$ for the same samples.\label{vavsaxs}}
\end{figure}

\begin{figure}

\caption{a) Sound velocity as a function of temperature for 6 float 
glass samples prepared with different fictive temperatures. b)
 Brillouin shift (proportional to sound velocity) at room temperature
 as   a function of $T_f$. The error bar on sound velocity are 0.1 \%. \label{vavbril}}
\end{figure}

\begin{figure}

\caption{Evolution of $I(q=0)$, with time during annealing 
\label{silicerelax}.  Insert:  $I(q=0)$, as a function of 
temperature for as-received sample during temperature 
dependent measurements (crosses), for as received sample  
during annealing at 1175 $^{\circ}$C (triangles), and 
for sample of \mbox{$T_f=1200$ $^{\circ}$C} (circles). 
The error on $I(q=0)$ is $\pm 2\%$}
\end{figure}

\newpage
\epsfxsize=230pt{\epsffile{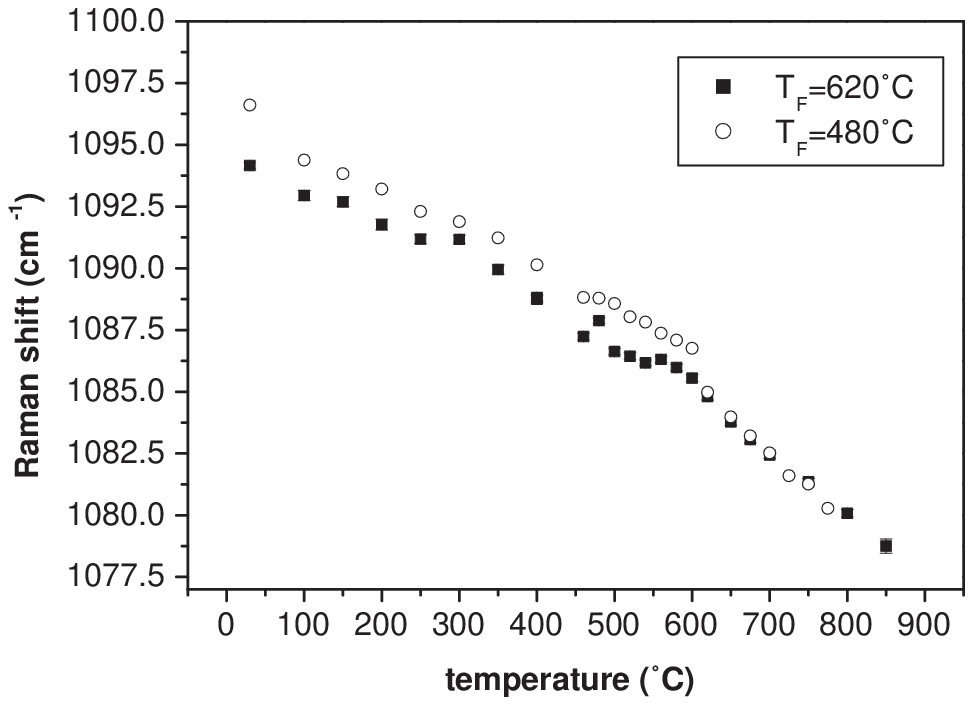}}\epsfxsize=220pt{\epsffile{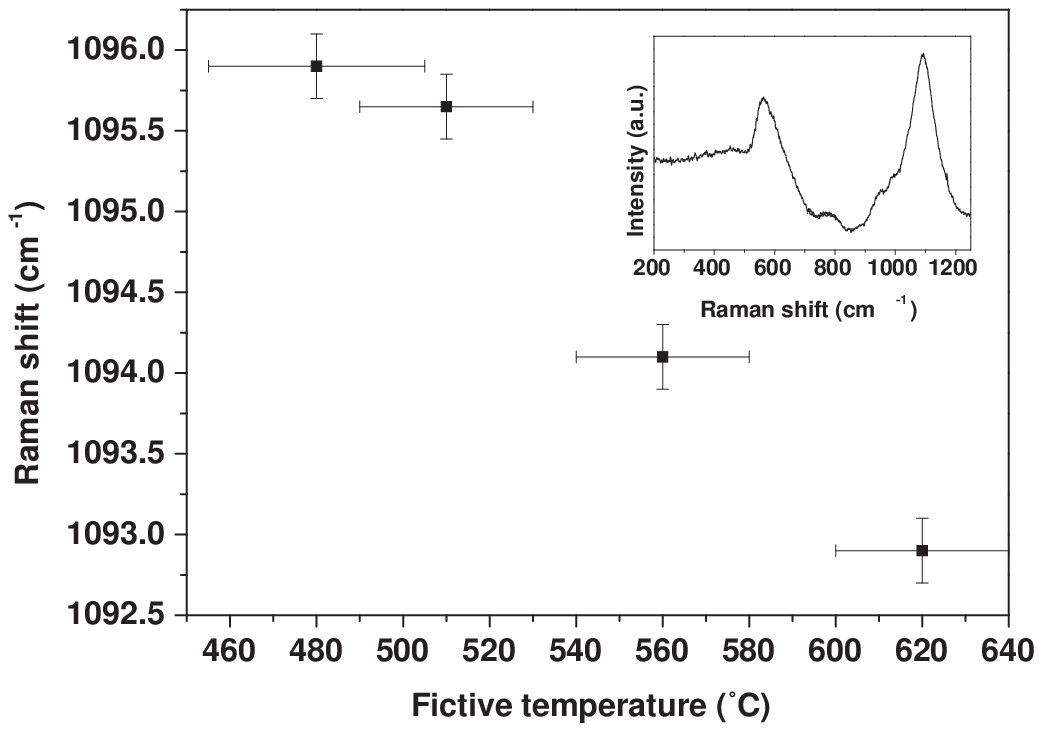}}
Fig. \ref{vavram}
\newpage 
\epsfxsize=240pt{\epsffile{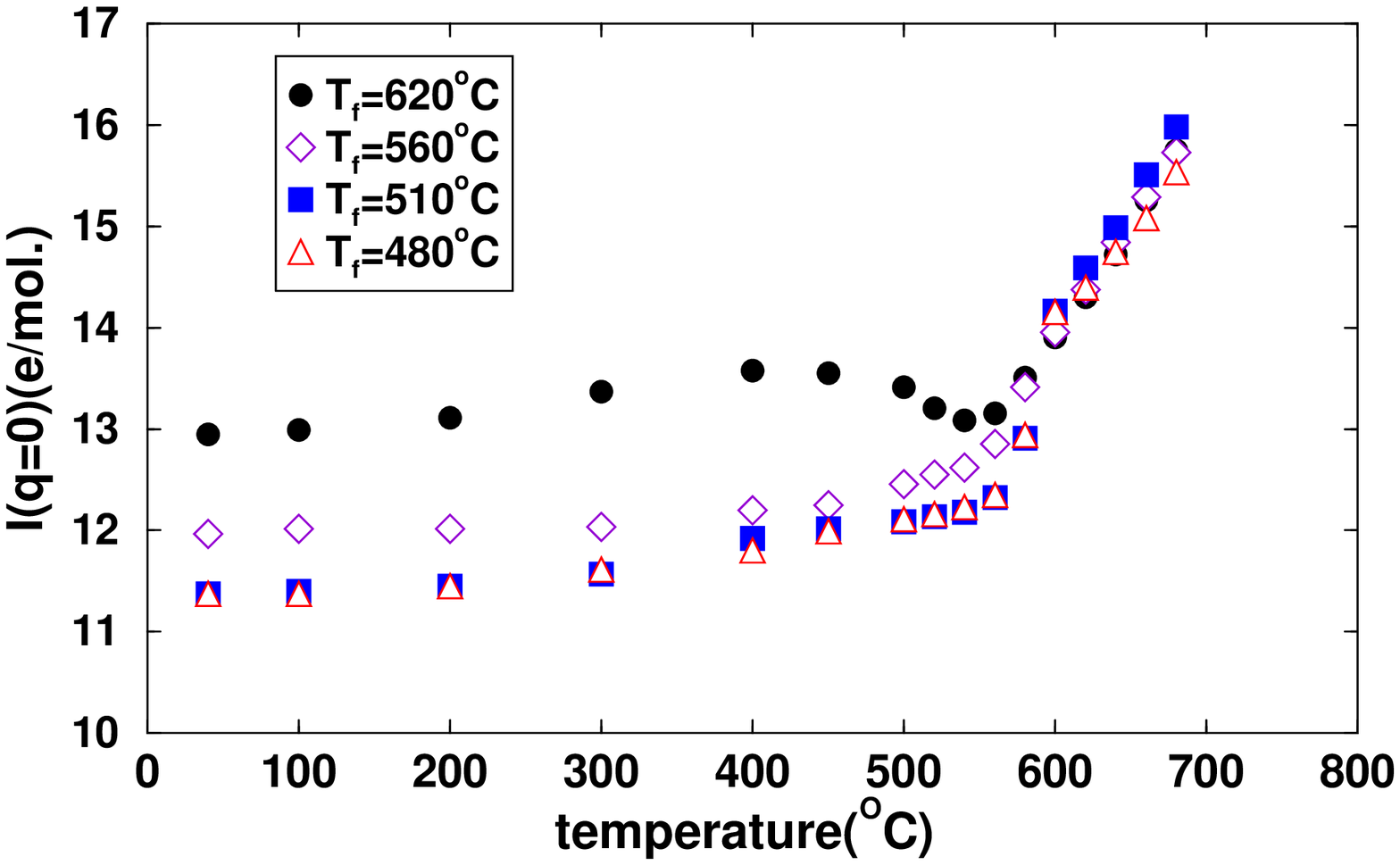}}\epsfxsize=190pt{\epsffile{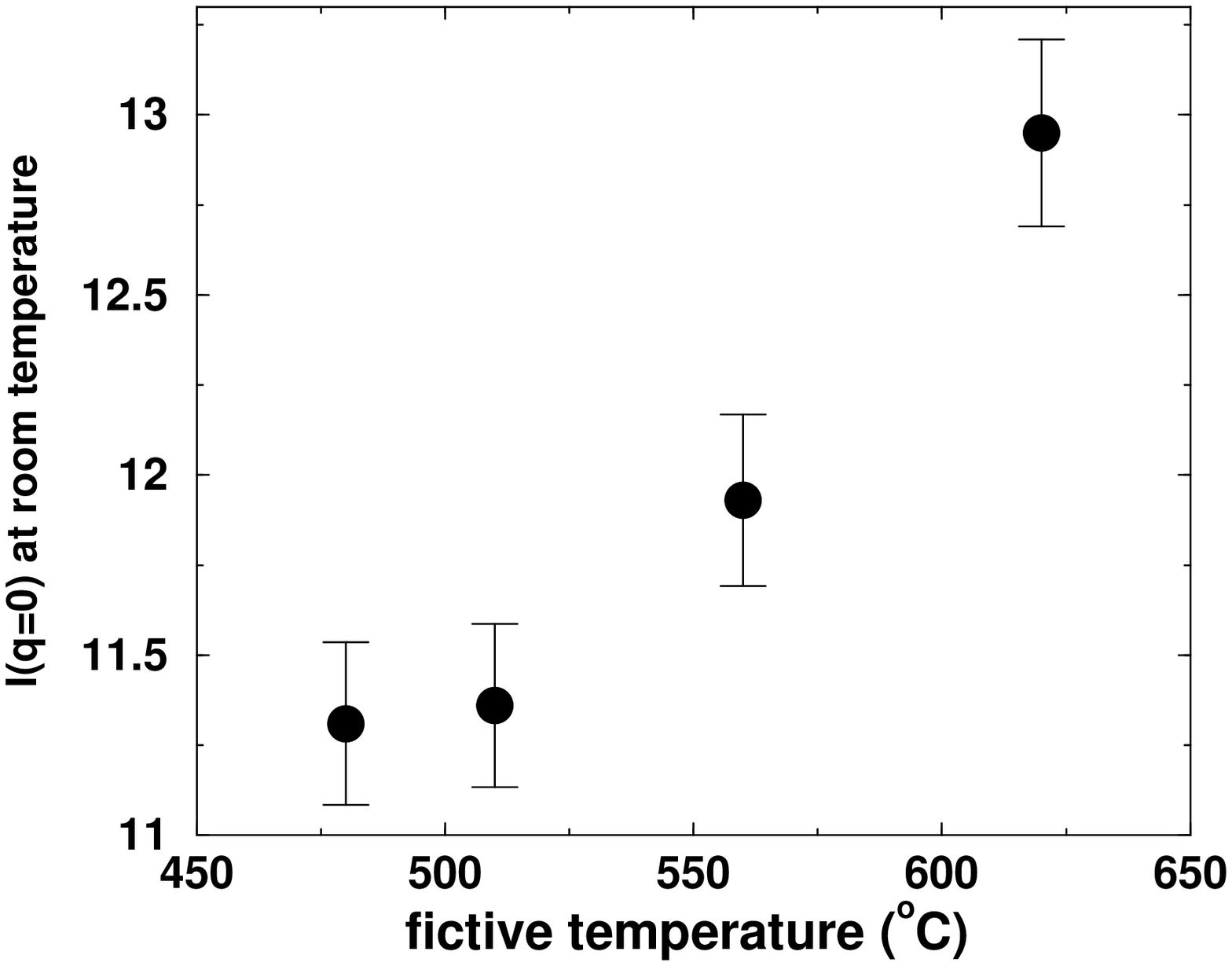}}

Fig. \ref{vavsaxs}
\newpage
\epsfxsize=210pt{\epsffile{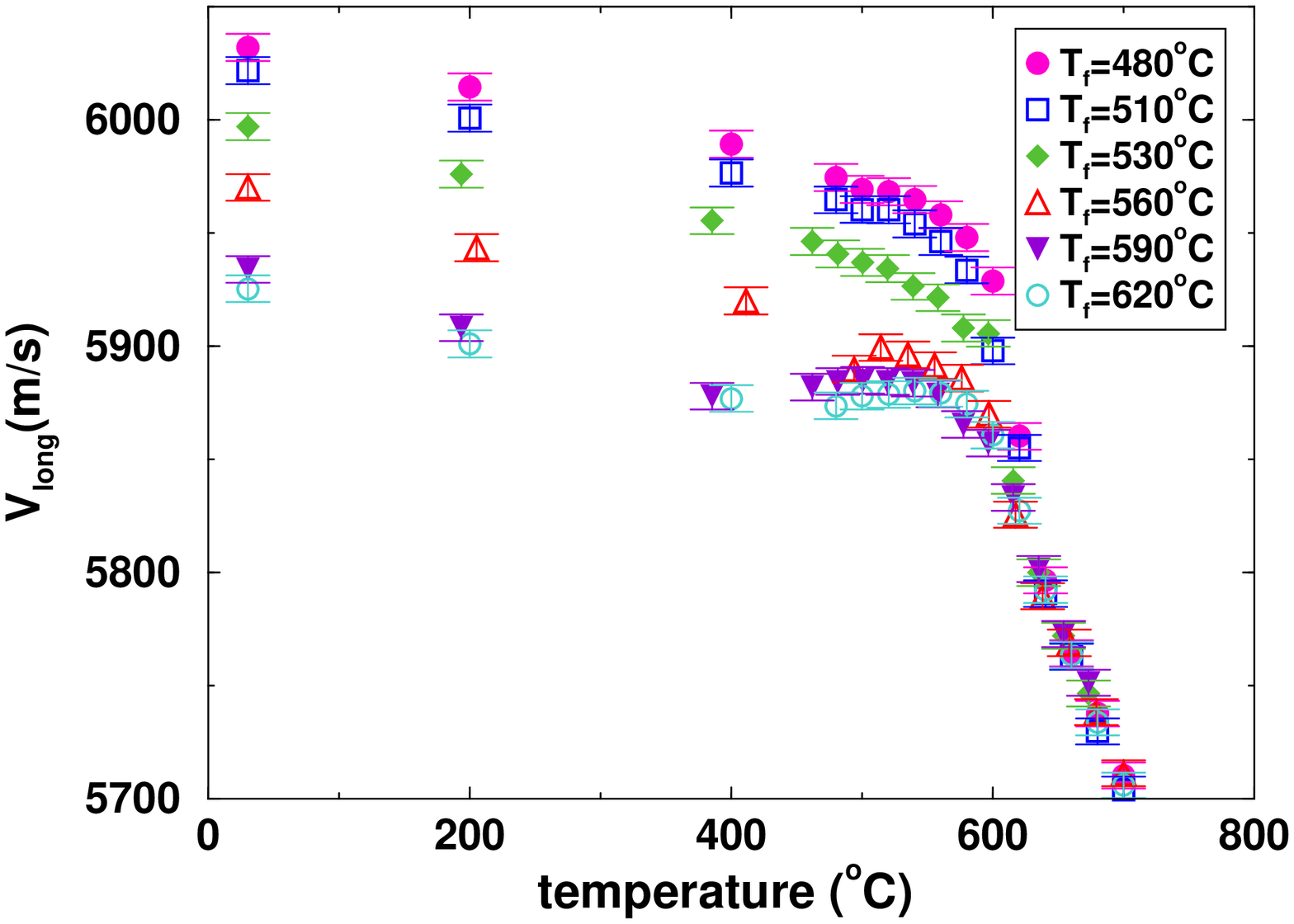}}\epsfxsize=210pt{\epsffile{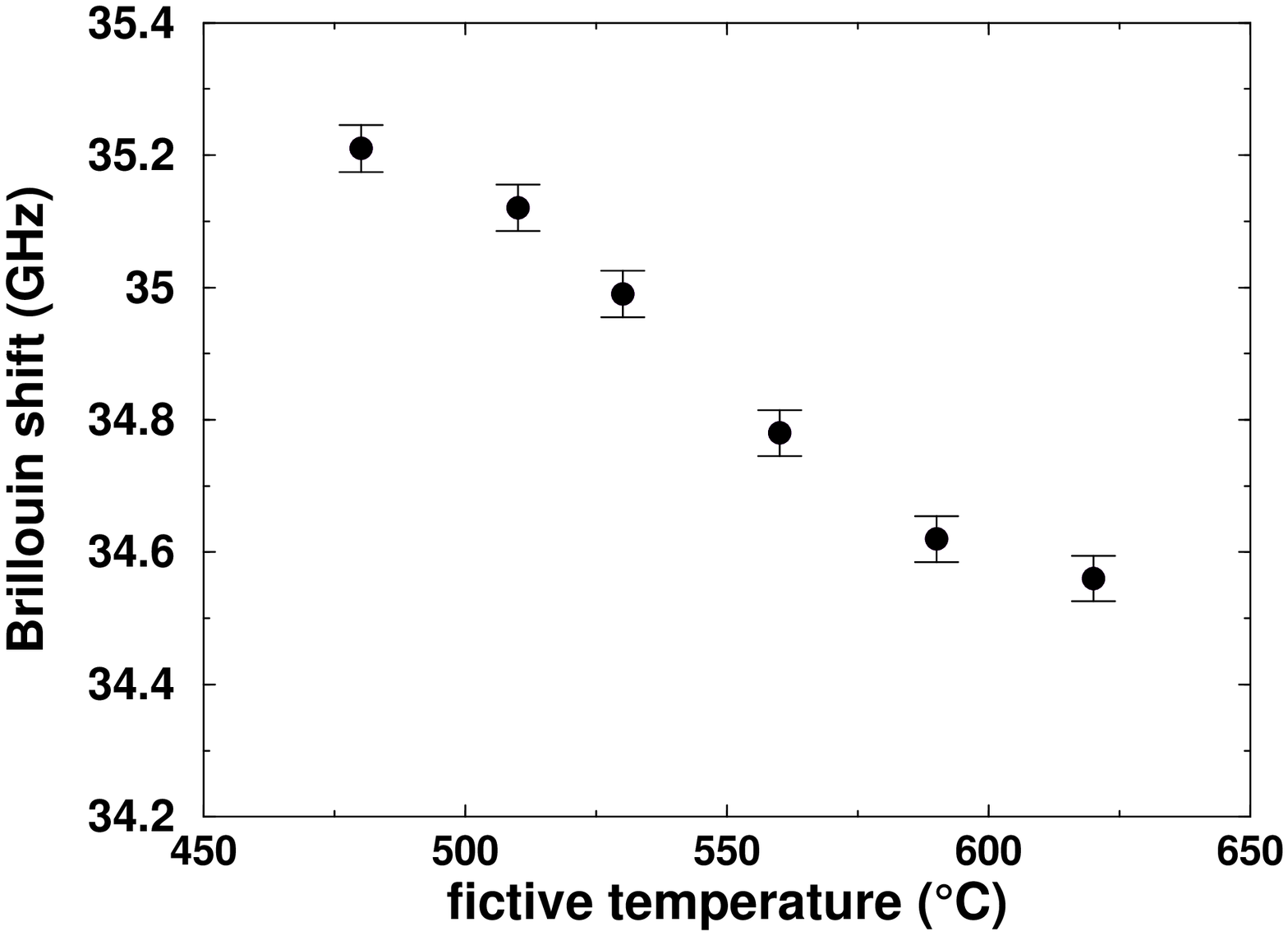}} 
Fig. \ref{vavbril} 

\newpage
\centerline{\epsfxsize=260pt{\epsffile{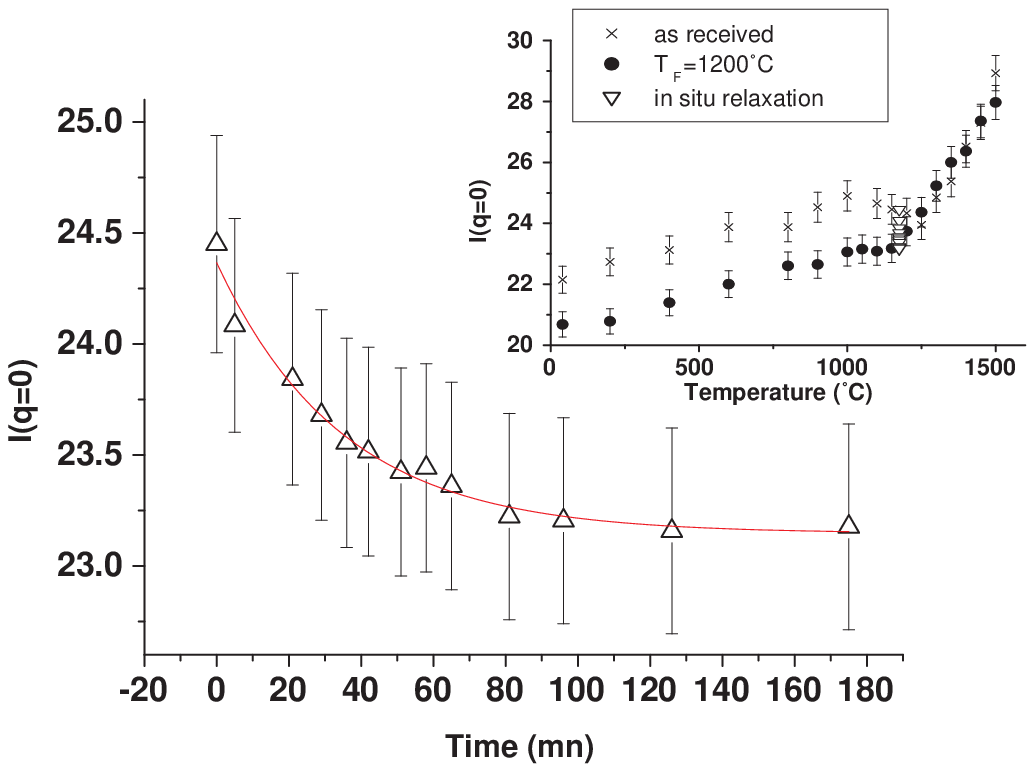}}}
Fig. \ref{silicerelax}
\end{document}